\documentclass[aps,prb,twocolumn,superscriptaddress,showpacs]{revtex4}
\usepackage{graphicx}

\begin{document}

\title{Magnetic field dependence of hole levels in self-assembled InAs quantum dots}

\author{J.I. Climente}
\affiliation{CNR-INFM-S3, Universit\`a degli Studi di Modena e Reggio Emilia, Via Campi 213/A, 41100 Modena, Italy}
\affiliation{Departament de Ci\`encies Experimentals, Universitat Jaume I, Box 224, E-12080 Castell\'o, Spain}
\author{J. Planelles}
\email{planelle@exp.uji.es}
\affiliation{Departament de Ci\`encies Experimentals, Universitat Jaume I, Box 224, E-12080 Castell\'o, Spain}
\author{M. Pi}
\affiliation{Departament d'Estructura i Constituients de la Mat\`eria, Universitat de Barcelona, E-08028 Barcelona, Spain}
\author{F. Malet}
\affiliation{Departament d'Estructura i Constituients de la Mat\`eria, Universitat de Barcelona, E-08028 Barcelona, Spain}

\date{\today}

\begin{abstract}

Recent magneto-transport experiments of holes in InGaAs quantum dots
[D. Reuter, P. Kailuweit, A.D. Wieck, U. Zeitler, O. Wibbelhoff, C. Meier,
A. Lorke, and J.C. Maan, Phys. Rev. Lett. {\bf 94}, 026808 (2005)]
are interpreted by employing a multi-band ${\mathbf k} \cdot \mathbf{p}$ Hamiltonian, which considers the interaction
between heavy hole and light hole subbands explicitely.
No need of invoking an incomplete energy shell filling is required within this model.
The crucial role we ascribe to the heavy hole-light hole interaction is further
supported by one-band local-spin-density functional calculations, which show that
Coulomb interactions do not induce any incomplete hole shell filling and therefore
cannot account for the experimental magnetic field dispersion.

\end{abstract}


\maketitle


Knowledge of the electronic structure of semiconductor quantum dots (QDs)\cite{Jacak_book} is essential
to understand and predict their physical properties.
The energy structure of conduction band electrons confined in QDs has been thoroughly investigated
using capacitance-voltage (C-V)\cite{WarburtonPRB} and far-infrared absorption spectroscopy.\cite{FrickeEL}
However, direct and clear access to the energy structure of valence band holes has long been hindered
by the high density of states arising from their larger effective mass.\cite{MedeirosAPL;SauvageAPL}
Only very recent C-V spectroscopy experiments have succeeded in resolving a number of energy levels 
corresponding to holes confined in In(Ga)As self-assembled QDs.\cite{BockAPL,ReuterPE04,ReuterPRL,ReuterPE}
Of particular interest are the results reported by Reuter et al.,\cite{ReuterPRL,ReuterPE}
which studied the energy dispersion of hole charging peaks in QDs
with a magnetic field $B$ applied along the growth direction.
It was found that, for QDs charged on average with one or two holes, the slope of the 
peaks dispersion was approximately zero. When the QDs were charged with three to six holes, the peaks
shifted towards lower and higher energy in alternating fashion, the slope of the fifth and sixth
peaks being about twice as strong as that of the third and fourth ones.
By using a one-band effective mass Hamiltonian, such results were interpreted as a non-sequential shell filling,
 leading to highly spin-polarized ground states.\cite{ReuterPRL}
It was speculated that this phenomenon, which has no analogy in conduction band electrons, might be due to the
Coulomb exchange interactions exceeding the rather small kinetic energy of holes.

In this paper, we provide theoretical support for these experiments. First, we study the effect of Coulomb
correlations on the electronic configuration of the hole ground state in self-assembled In(Ga)As QDs.
To this end, we use a one-band local-spin-density functional (LSDF) model. We find that, even for a rather large QD and
pure heavy hole effective mass, a sequential filling of the energy levels is predicted.
It then follows that an explanation for the experimental magnetic field-induced dispersion of the charging peaks 
is unattainable within the one-band picture. 
Next, we study the role of heavy hole-light hole (HH-LH) coupling, using a four-band
Luttinger-Kohn Hamiltonian.  Our results show that HH-LH interaction, together with the spin Zeeman
splitting, yield a plausible interpretation of the experimental data even at an independent particle level.


Structural information of self-assembled QDs is difficult to assess, and this imposes a severe restriction 
for a quantitative description of their physics.
Nonetheless, the sample of QDs studied in Refs.\onlinecite{ReuterPRL,ReuterPE} showed a narrow size distribution,
and a rough guess of their shape, size and composition is possible. 
Comparing the growth parameters and the photoluminescence emission wavelength to other experiments and theoretical work, 
the QDs are believed to be lens shaped, the base plane being approximately circular but slightly elongated along 
the [0,1,1] direction.
The QDs height after overgrowth is believed to be between 4 nm and 6 nm, and the diameter between 32 nm and 40 nm.
The QDs are not pure InAs but have an inhomogeneous InGaAs composition.\cite{ReuterPC}

In order to see whether Coulomb correlations can induce an incomplete shell filling in such structures,
we investigate the electronic configuration and addition energy of the $N$-hole ground state at zero magnetic field
($N$=1-6) using an effective mass LSDF model.\cite{PiPRB} 
The theoretical model is fully three-dimensional, which allows us to represent the QDs as circularly symmetric 
lenses with finite height, and use bulk material parameters without any need to fit them in order to account 
for the missing vertical motion.\cite{PeetersPRB}
We probe two model QDs, which constitute a lower and an upper estimate of the possible experimental sizes. 
The smaller (larger) QD has a radius of $R=16$ (20) nm, and a height of $h=4$ (6) nm. 
In both cases, we take an average effective mass value between that of InAs and that of GaAs heavy holes, 
$m^*=0.44$.\cite{Landolt_book}
The confinement potential is defined by a step-like function whose height is given by the InGaAs valence band-offset, 
380 meV,\cite{VurgaftmanJAP} and the effective dielectric constant we use is $12.4$. 

Figure 1 represents the addition spectra, $\Delta A(N) \equiv E(N+1)-2\,E(N)+E(N-1)$, where
$E(N)$ is the energy of the $N$-hole ground state, for the  smaller (solid line) and larger (dashed line) QDs.
The insets show the occupation of the Kohn-Sham orbitals for each $N$, the arrows representing the sign of the spin.
It can be seen that the spectra of both QDs exhibit the typical behavior found for conduction band electrons confined 
in axially symmetric QDs,\cite{PiPRB,TaruchaPRL}
with local maxima at shell filling values ($N=2,6$) and at half-filling values ($N=4$).
These results are consistent with a sequential filling of the hole energy shell, according to the Aufbau principle and Hund's rule.
It is worth noting that this result holds in spite of the fact that our model QDs are more correlated than the
experimental ones.  For instance, the Coulomb blockade energy between the two first holes in the experiment is
$24$ meV,
similar to the value of $20.2$ meV we obtain for the smaller QD. In contrast, 
the kinetic energy of our calculations is fairly smaller (about one third) than that inferred from the 
experimental data, in part owing to the assumption of pure heavy hole behavior. 
Even for the larger QD,
no strongly spin-polarized system is predicted.
At this point, one can notice that filling the hole energy shells sequentially means that
the fifth and sixth holes occupy orbitals with the same absolute value of the envelope angular momentum 
z-projection as the third and fourth ones,
namely $|m_z|=1$. Therefore, the experimental observation that the magnetic field 
dispersion of the $N$=5 and $N$=6-hole states is twice as strong as that of the $N$=3 and $N$=4 ones
can no loger be explained in terms of different angular momenta within the one-band model.
However, different angular momenta may still be possible within a multi-band model, as HH-LH interaction
yields hole states with different mixtures of envelope angular momenta, which has a dramatic effect
on the magnetic field dispersion.\cite{RegoPRB}
We thus proceed to investigate role of HH-LH interaction in our model QDs.

We use a three-dimensional, four-band ${\mathbf k} \cdot {\mathbf p}$ Hamiltonian in the envelope function approximation,
with a magnetic field applied along the growth direction.\cite{ClimentePRBeh}
The QD we investigate has the same dimensions and confinement potential as the smaller dot of previous paragraphs.
We use InGaAs Luttinger parameters ($\gamma_1$=$11.01$, $\gamma_2$=$4.18$, $\gamma_3$=$4.84$),\cite{VurgaftmanJAP}
which yield effective masses of $m^*_{HH}$=$0.377$ and $m^*_{LH}$=$0.052$.
Figure 2 (solid lines) shows the lowest-lying single-particle energy levels as a function of the magnetic field. 
The levels are labeled by their total angular momentum $z$-projection, $F_z=m_z+J_z$, where 
$m_z=0,\,\pm 1,\,\pm 2\ldots$ and $J_z=\pm 3/2,\, \pm 1/2$ are the envelope and Bloch angular momentum
$z$-projections, respectively.
The slopes of the energy levels vs.\ $B$ are the result of the particular mixture of envelope angular 
momenta for each level (see Table I).
For example, the predominant component of the $F_z=3/2$ state at $B=0$ is that of $J_z=3/2$ 
(about $97.5$\% of the charge density), which is associated with $m_z=0$. 
This component is insensitive to the linear term of the magnetic field (orbital Zeeman splitting).\footnote{All the
low-lying states of our model QDs have negligible diamagnetic shifts in the magnetic field range under study.}
However, there is also a small contribution ($2.5$\%) coming from the components associated with $m_z=1,\,2,\,3$,
which introduces a slight dependence on the magnetic field.
Likewise, for $F_z=1/2$ the leading component (87\%) is $J_z=3/2$, which is
associated with $m_z=-1$. However, a non-neglegible $J_z=-1/2$ component (7\%) with $m_z=1$ partially
compensates for the $m_z=-1$ magnetic field dispersion, so that the final slope is not as steep as it would be
in a one-band model.
Finally, the $F_z=5/2$ state has a predominant (93\%) $J_z=3/2$ component, connected with $m_z=1$, 
and minor components with $m_z=2,\,3,\,4$, which enhance the magnetic field dispersion as compared to the
one-band model.
Similar reasonings can be made for $F_z=-3/2,\,-1/2$ and $-5/2$ states.
For a more accurate description of the magnetic field dispersion, we need to add the spin Zeeman 
term, $\kappa\,\mu_B\,B\,J_z$, on the diagonal elements of our four-band Hamiltonian. In this term, $\mu_B$ 
is the Bohr magneton and $\kappa$ the Land\'e factor for holes, which we take close to its value 
for GaAs, $\kappa=1.2$.\footnote{The Land\'e factors for
InAs and GaAs are $\kappa=7.68$ and $\kappa=1.2$, respectively. However, these factors are
very sensitive to the semiconductor composition, band mixing, spatial confinement and 
strain,\cite{SnellingPRBNakaokaPRB}
so that an \emph{ab initio} value for our QDs is hard to justify. Nevertheless, we note that the $k$ we use is of a
similar order of magnitude to that inferred by Bayer et al\ for a sample of In$_{0.60}$Ga$_{0.40}$As QDs
($\kappa=0.73$).\cite{BayerPRL} In this way, the spin Zeeman shifts we predict up to 10 T are about $0.8$ meV, 
a bit larger than those expected from Ref.\onlinecite{ReuterPRL} experiments with a magnetic field 
applied parallel to the plane of the QDs ($1.5$ meV up to 28 meV).
This difference may be explained in terms of the Land\'e factor anisotropy.}
It is important to note that the spin Zeeman term, often neglegible in the description of conduction band electrons,
may become comparable to the orbital Zeeman term for low-lying hole states. 
This is because such states have strong heavy hole character and thus a rather large effective mass 
acting upon the linear term of the magnetic field.
The result of adding the spin Zeeman energy is illustrated in Figure 2 (dashed lines).
It can be seen that the magnetic field dispersion of the $|F_z|=3/2$ and $|F_z|=5/2$ states is slightly
quenched, while that of the $|F_z|=1/2$ states is enhanced.
The underlying reason is that the signs of $J_z$ and $m_z$ in the leading component of the $|F_z|=1/2$ states are opposite.

Next, we calculate the energy of the charging peaks vs.\ $B$ which would follow from filling the energy levels
of Figure 2 in an independent particle scheme, $\mu(N,B)=E(N,B)-E(N-1,B)$.
To compare the different slopes of the individual charging peaks, we substract the peak energy at $B=0$
and represent the absolute value in Figure 3.
By comparing the figure to Figure 3 of Ref.\onlinecite{ReuterPRL} in the same range of magnetic
field ($B$=0-10 T), a fairly good agreement is observed.
For example, the average slope of the first and second peaks is $0.04$ meV/T (close to the experimental value
of $0.03$ meV/T), the average slope of the third and fourth peaks is $0.18$ meV/T ($0.14$ meV/T in the experiment),
and the average slope of the fifth and sixth peaks is $0.32$ meV/T ($0.33$ meV/T in the experiment). 
Our results therefore suggest that it is HH-LH interaction rather than Coulomb interaction the key factor on the
experimentally observed behavior, although of course the latter may help to improve the quantitative
description of the system.
Finally, we would like to point out that for QDs with other dimensions than those of the dot we have investigated
here, the order of the excited single-particle energy levels may differ.
However, agreement with the experimental results is only achieved for the energy level sequence shown in 
Figure 2.
We would like to note though that a very recent theoretical work also investigating the hole electronic configuration
of self-assembled InAs QDs, albeit in absence of magnetic field, [L. He, G. Bester and A. Zunger,
arXiv:cond-mat/0505330] points out that Coulomb interaction induces a change in the hole shell filling sequence.
However, the QDs they study are significantly smaller than those expected from Ref.\onlinecite{ReuterPC},
and in any case the resulting electronic configuration upon inclusion of the Coulomb term is equivalent to that
we find in Figure 2 at an independent particle level.


We thank M. Barranco, D. Reuter and P. Kailuweit for many useful suggestions.
Financial support from MEC-DGI project CTQ2004-02315/BQU and UJI-Bancaixa project P1-B2002-01 (JP,JIC),
BFM2002-01868  (DGI, Spain) and 2001SGR00064 (Generalitat de Catalunya) (MP,FM)
are gratefully acknowledged.
This work has been supported in part by the EU under the TMR network ``Exciting'' (JIC).



\begin{figure}[p]
\includegraphics[width=0.8\textwidth]{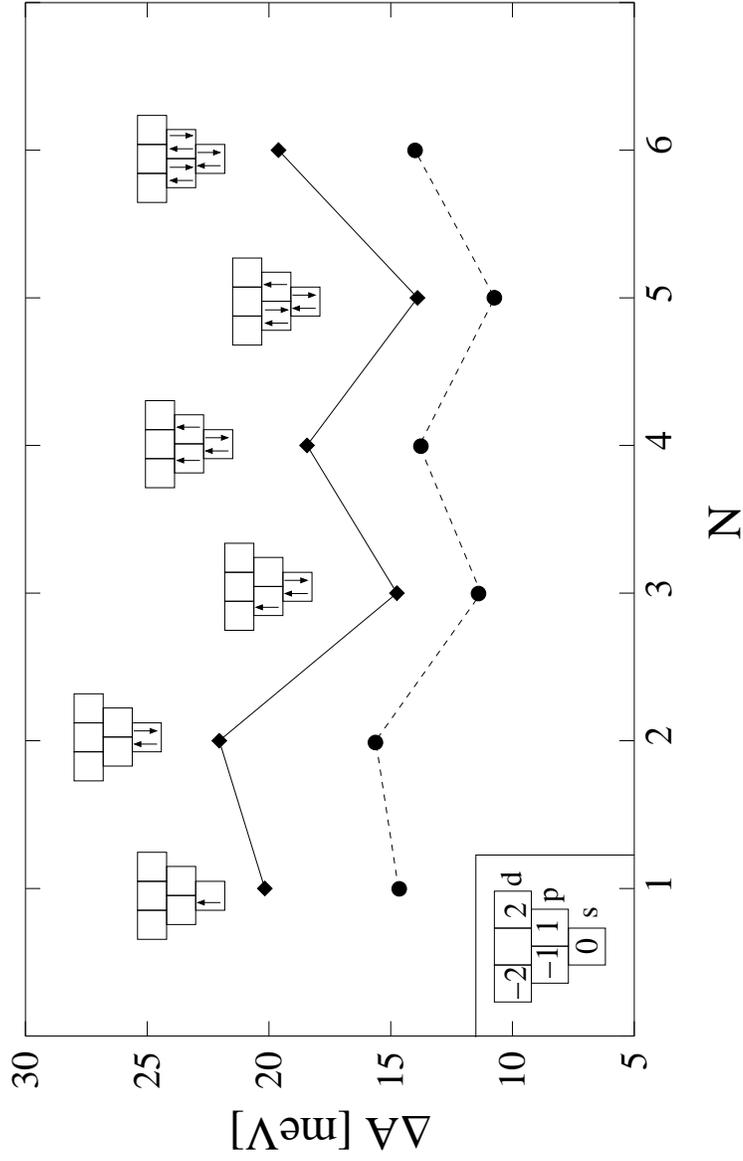}
\caption{Addition energy vs.\ number of holes in a QD with $R=16$ nm and $h=4$ nm (solid line)
and in a QD with $R=20$ nm and $h=6$ nm (dashed line). The insets show the occupation of the
Kohn-Sham orbitals for each $N$, the arrows representing the sign of the spin.}
\label{Fig1}
\end{figure}

\begin{figure}[p]
\includegraphics[width=0.8\textwidth]{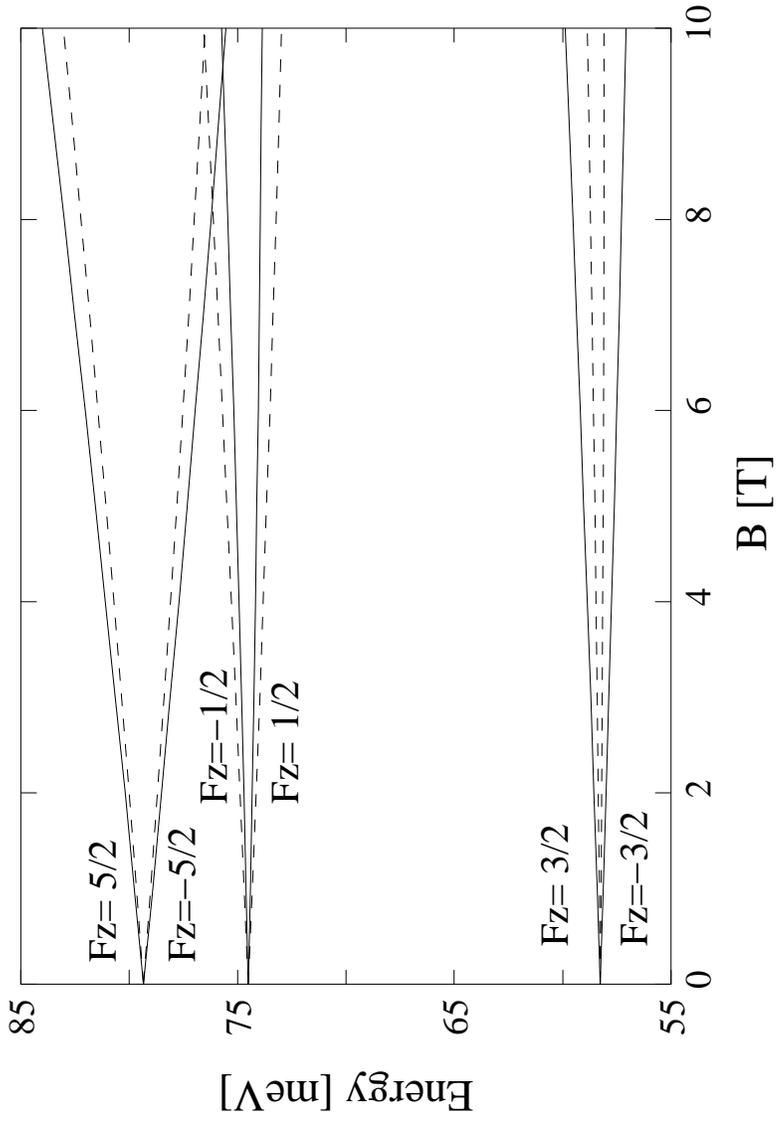}
\caption{Single-particle, four-band energy levels vs. magnetic field for a QD with $R=16$ nm and $h=4$ nm.
Solid (dashed) lines are used for the energy levels calculated without (with) spin Zeeman splitting.}
\label{Fig2}
\end{figure}

\begin{table}[p]
\begin{tabular}{|l|c|c|c|c|}
\hline
                & $J_z=3/2$	& $J_z=1/2$	& $J_z=-1/2$	& $J_z=-3/2$	\\
\hline
$F_z=3/2$ \begin{tabular}{|c} $m_z$ \\ weight \end{tabular}
		&
	  \begin{tabular}{c} 0 \\ 0.975 \end{tabular}
		&
	  \begin{tabular}{c} 1 \\ 0.018 \end{tabular}
		&
	  \begin{tabular}{c} 2 \\ 0.006 \end{tabular}
		&
	  \begin{tabular}{c} 3 \\ 0.001 \end{tabular} \\
\hline
$F_z=1/2$ \begin{tabular}{|c} $m_z$ \\ weight \end{tabular}     
		&
	  \begin{tabular}{c} -1 \\ 0.868 \end{tabular} 
		&
	  \begin{tabular}{c}  0 \\ 0.047 \end{tabular} 
		&
	  \begin{tabular}{c}  1 \\ 0.072 \end{tabular}
		&
	  \begin{tabular}{c}  2 \\ 0.013 \end{tabular} \\
\hline
$F_z=5/2$ \begin{tabular}{|c} $m_z$ \\ weight \end{tabular}
		&
	  \begin{tabular}{c}  1 \\ 0.928 \end{tabular}
		&
	  \begin{tabular}{c}  2 \\ 0.041 \end{tabular}
		&
	  \begin{tabular}{c}  3 \\ 0.029 \end{tabular}
		&
	  \begin{tabular}{c}  4 \\ 0.002 \end{tabular} \\
\hline
\end{tabular}
\label{Tab1}
\caption{Envelope angular momentum z-projections and charge density weights associated with each $J_z$ component
of the three lowest-lying single-particle hole states at $B=0$. Only the states with positive $F_z$
are shown. States with negative $F_z$ have reversed order of weights and $m_z$, and the sign of $m_z$ changes.} 
\end{table}

\begin{figure}[p]
\includegraphics[width=0.8\textwidth]{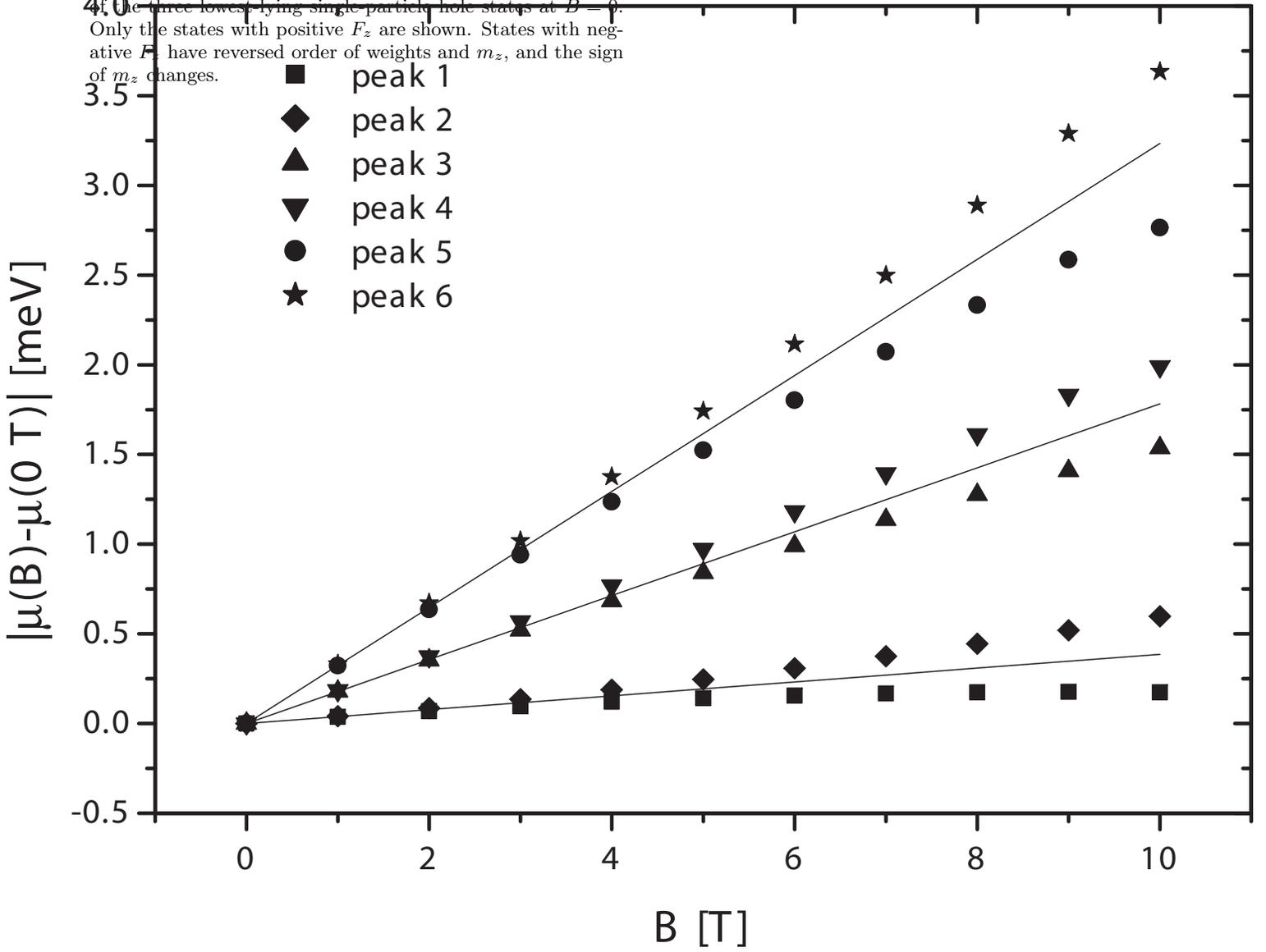}
\caption{Absolute value of the magnetic field-induced energy shift for charging peaks of 1 to 6 holes
vs.\ the magnetic field. Solid lines represent the average slope of each pair of peaks.}
\label{Fig3}
\end{figure}

\end{document}